\pdfoutput=1
\documentclass{aa}  

\usepackage[normalem]{ulem}

\usepackage{graphicx}
\usepackage{txfonts}
\usepackage{slashbox}
\usepackage{mhchem}
\usepackage{xcolor}
\usepackage[normalem]{ulem}

\definecolor{deepmagenta}{rgb}{0.8, 0.0, 0.8}

\usepackage{afterpage}

\usepackage[colorlinks=true,linkcolor=blue,citecolor=blue,urlcolor=blue]{hyperref}

\newcommand{\cair} {
  \ion{Ca}{ii}~8542~\AA\:}
  
\newcommand{\halpha} {
  \ce{H\alpha}}

\newcommand{\cak} {
  \ion{Ca}{ii}~K\:}
\begin{document}

\newcommand{\com}[1]{{\color{orange}{#1}}}

   \title{Physical properties of a fan-shaped jet backlit by an X9.3 flare}

   \author{A.G.M. Pietrow\inst{1}
   \and M. Druett \inst{1}
   \and J. de la Cruz Rodriguez\inst{1}
   \and F. Calvo\inst{1}
   \and D. Kiselman\inst{1}
          }

   \institute{Institute for Solar Physics, Dept. of Astronomy, Stockholm University, Albanova University Centre, SE-106 91 Stockholm, Sweden\\
              \email{alex.pietrow@astro.su.se}
             }

   \date{Received November 1, 2021; accepted November 25, 2021}

\abstract{Fan-shaped jets sometimes form above light bridges and are believed to be driven by reconnection of the vertical umbral field with the more horizontal field above the light bridges. Because these jets are not fully opaque in the wings of most chromospheric lines, one cannot study their spectra without the highly complex considerations of radiative transfer in spectral lines from the atmosphere behind the fan.}
{We take advantage of a unique set of observations of the \halpha ~line along with the \cair and \cak lines obtained with the CRISP and CHROMIS instrument of the Swedish 1-m Solar Telescope to study the physical properties of a fan-shaped jet that was backlit by an X9.3 flare. For what we believe to be the first time, we report an observationally derived estimate of the mass and density of material in a fan-shaped jet. }
{The \halpha~ flare ribbon emission profiles from behind the fan are highly broadened and flattened, allowing us to investigate the fan with a single slab via Beckers' cloud model, as if it were backlit by continuous emission. Using this model we derived the opacity and velocity of the material in the jet. Using inversions of \cair emission via STiC (STockholm inversion Code), we were also able to estimate the temperature and cross-check the velocity of the material in the jet. Finally, we use the masses, plane-of-sky (POS) and line-of-sight velocities as functions of time to investigate the downward supply of energy and momentum to the photosphere in the collapse of this jet, and evaluate it as a potential driver for a sunquake beneath.}
{We find that the physical properties of the fan material are reasonably chromospheric in nature, with a temperature of 7050 $\pm$ 250 K and a mean density of 2 $\pm$ 0.3 $\times$ 10$^{-11}$ g cm$^{-3}$.}
{The total mass observed in \halpha~ was found to be 3.9 $\pm$ 0.7 $\times$ 10$^{13}$g and the kinetic energy delivered to the base of the fan in its collapse was nearly two orders of magnitude below typical sunquake energies. We therefore rule out this jet as the sunquake driver, but cannot completely rule out larger fan jets as potential drivers.}
 

   \maketitle
%

\section{Introduction}

\halpha \: surges were first reported in \citet{1937McMath}, where they were classified as class IIId prominences. Since then surges have been a point of interest and ongoing solar research \citep[e.g.][]{1942newton, 1949ellison, 1961malville, 1994schmeieder, 2012Uddin, 2018carolina, 2021nobrega}.

These structures are described as straight or slightly curved spikes of cool (<15kK) material that appear most often in absorption in chromospheric lines such as \halpha~ and \cair~ on the solar disk \citep{1977bruzek}. They originate from small bright points close to sunspots and pores, and are believed to be the result of reconnection events between newly emerging flux and a preexisting magnetic field \citep[e.g.][]{1995yokoyama, 2013Kayshap, carolina16}. However, it has also been shown that they can be triggered by the impulsive generation of a pressure pulse \citep{1993Sterling}, and be associated with explosive events \citep{2009Madjarska}. Typical lifetimes for surges are between 10 and 20 minutes, but longer lived surges of over 40 minutes and even over an hour have been reported \citep[e.g.][]{2018carolina, 2021Kayshap}. Jets studied by \citet{1973royPhd, 1973royb, 1973roy} reached heights between 7 and 50 Mm, but much higher jets up to 200 Mm have been reported \citep[e.g.][]{1977bruzek,2021Louis}. The material that is propelled upwards in surges is thought principally to return approximately along the trajectory of ascent, with some small fraction possibly being heated to coronal values and remaining at higher locations in the solar atmosphere \citep{carolina16, 2018carolina}. In \citet{1973roy} it was reported that the material in surges accelerated more slowly than the effective gravity (the solar gravitiational component along the trajectory of ascent and descent), suggesting the presence of a breaking force. However this effect was not found in more recent work by \citet{carolina16}, where the acceleration was found to be consistent with effective solar gravity.

\citet{1977bruzek} suggest that the magnetic field strength inside surges is around 50~G and decreases with height, which is consistent with by \citet{1969Harvey}, who measured field strengths between 17 and 87~G in surges on the limb. A more recent estimate was made by \citet{2003Yoshimura}, who suggested a field strength between 100 and 300~G based on an estimate of the reconnection energy. The models of \citet{2010Ding} produced still lower field strength values, between 6 and 40~G. \citet{2010Ding} found that in their models the jet temperature depended on the magnetic field strength and the jet density. A jet with a weak magnetic field (6G) and high density showed lower, more chromospheric, temperatures ($\approx 10^4$~K), while a jet with a higher magnetic field (40G) and lower density would result in a hot surge with much higher temperatures of around $6\times10^5$~K. Simulations producing 'hot surges' were also reported in e.g. \citet[][]{2008Nishizuka, 2013Kayshap}. Hot surge simulation results have been challenged more recently in \citet{2016nobrega}, who suggest that hot surges will likely cool down rapidly to $\approx 10^4$~K when detailed radiative losses are considered. A surge with lower temperature have also been reported more recently, with observed brightness temperature of 5.5~kK by \citet{carolina16}.

Fan-shaped jets, or peacock jets, are a subclass of surges that appear over light bridges. This was first reported by \citet{1973roy, 1973royb} and studied as an independent phenomenon by \citet{2001asai}, where the jets were described as recurring jets that appear dark in \halpha~ and are also visible in the extreme ultra~violet. A measurement of the line-of-sight (LOS) magnetic field by \citet{2007bharti} shows that some of these jets can have a polarity opposite to that of the umbral field. Small jets up to 13Mm have been reported by \citet{2009Ashimizu2,2009shimizu} and much smaller jets, most of which no longer than 1000 km were reported by \citet{2014louis}.

Due to the complexity of the 3D radiative transfer problem and the highly variable background around sunspots, the inference of detailed physical parameters of the fan jets from inversions of the convolved spectra they produce in front of their backgrounds has so far proven to be an intractable task. This has limited many of the results that were presented above, especially with respect to temperature and density. As a result, there are no observationally constrained density measurements of surges, and typical chromospheric values are often assumed for simulations. An estimate of the mass and density of a large surge was made by \citet{1973roy} based solely on an isotropic expansion of chromospheric material from the base of the fan, giving values of order 10$^{15}$ and 10$^{16}$ g (density, 10$^{11}$ and 10$^{12}$ particles cm$^{-3}$). In this paper we will present a unique data set that allows us to make  what we believe to be the first observationally constrained mass and density measurement of a peacock jet. We also estimate the amount of returning mass that hits the base of the fan as a function of time. This in turn will allow us to explore the potential of surges as a trigger for sunquakes\footnote{Sunquakes were predicted \citet{1972wolf} and first observed by \citet{1998Kosovichev}.} by comparing the downward delivered momentum and energy budgets of material in the jet to that of measured sunquakes. We will not attempt to infer the magnetic field of the jet, given that these fields are weaker by several orders of magnitude than those of a flare \citep[e.g.][]{2021gragal}.

In Sect. \ref{observations} we discuss our observations, the reduction, and the spectral profiles found within. In Sect. \ref{datainversions} we discuss our models and the assumptions which allowed us to use these models.  The results from our inversions are shown and discussed in Sect. \ref{resultsndisc} and our final conclusions can be found in Sect. \ref{conclusions}.

\section{Observations and data processing}\label{observations}
Region AR12673 was observed on the 6th of September, 2017, between 11:55 and 12:52 UT with the Swedish Solar 1-m Telescope \citep[SST,][]{Scharmer03}, using both CRisp Imaging SpectroPolarimeter \citep[CRISP,][]{Scharmer08} and the CHROMospheric Imaging Spectrometer \citep[CHROMIS;][]{Scharmer17} simultaneously. The region was centred around the heliocentric coordinates  (x,y) = ($537\arcsec$, $-222\arcsec$), which translates to an observing angle of 37$^\circ$ ($\mu$ = 0.79). A sequence was run on CRISP that alternated between \halpha $\:$ and \cair, while on CHROMIS only the \cak line was observed. This dataset was first described in detail in \citet{Quinn19}, although the CHROMIS \cak data were reduced for the first time here, the \cak 4000\AA ~continuum point is employed for context images and to identify the location of the base of the fan jet that we investigated in this paper. In the relevant time frames the \cak line profiles did not have good enough seeing quality to significantly contribute to this project.

For CRISP the cadence was 15~s and the observing sequence consisted of 13 wavelength positions in the \halpha $\:$ line, at $\pm1.5$, $\pm1.0$, $\pm0.8$, $\pm 0.6$, $\pm 0.3$, $\pm 0.15$ and 0 \AA~ relative to line center. The \cair scan consisted of 11 wavelength positions taken in full-Stokes polarimetry mode at $\pm0.7$, $\pm0.5$, $\pm0.3$, $\pm 0.2$, $\pm 0.1$ and 0 \AA ~ relative to line center. The CRISP pixel scale is $0.058\arcsec$.

For CHROMIS the cadence of the observing sequence was 6.5~s with 10 wavelength positions in the \cak line at $\pm 1.00$, $\pm 0.85$, $\pm 0.65$, $\pm 0.55$, $\pm 0.45$, $\pm 0.35$, $\pm 0.25$, $\pm 0.15$, $\pm 0.07$ and 0.00 \AA~ relative to line center, plus a single continuum point at 4000~\AA~.  The CHROMIS pixel scale is $0.0375\arcsec$. In addition to the narrow band images, wide-band images were also obtained co-temporally with each CRISP and CHROMIS narrow band image, to be used for alignment purposes.

Both datasets were reduced again for use in this paper using the SSTRED pipeline as described by \citet{jaime15} and \citet{mats21}, which make use of Multi-Object Multi-Frame Blind Deconvolution \citep[MOMFBD;][]{vanNoort05,mats02}. Additionally, we performed an absolute wavelength and intensity calibration using the solar atlas by \cite{Neckel1984}. 

We extensively used the CRISPEX analysis tool \citep{Gregal12}, the CRISpy python package \citep{pietrow19} (Now part of the ISPy library \citep{ISPy2021}), COCOPLOTs \citep{2021druett} and SOAImage DS9 \citep{2003DS9} for data visualization during the preparation of this article. 

\subsection{Field of view}

\begin{figure*}[!htp]
   \centering
   \includegraphics[width=0.8\textwidth, trim=1.3cm 2cm 1.5cm 2cm,clip]{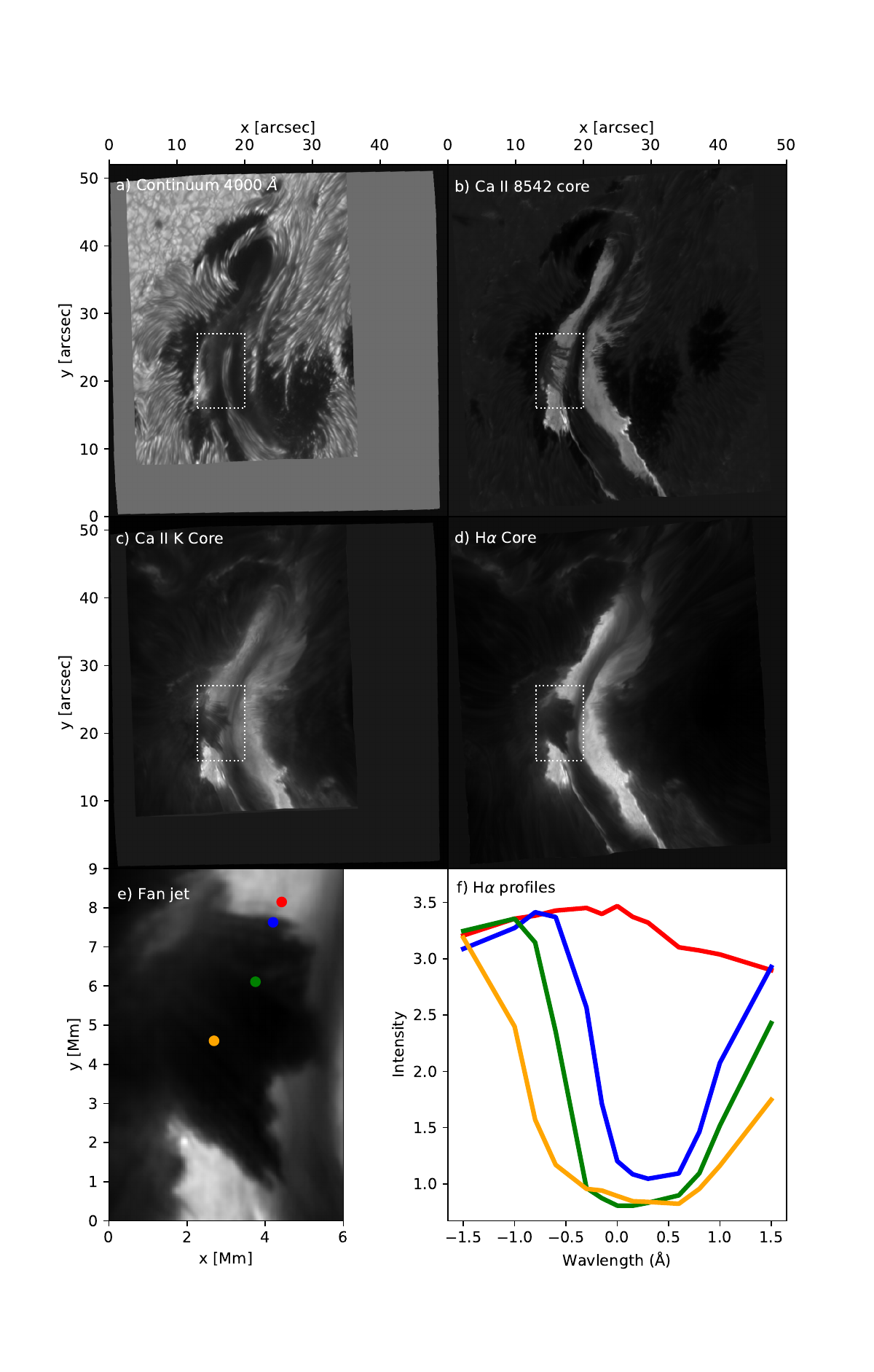}
   \caption{Overview of AR12673 taken on September 6th 2017 at 11:59 UT. The fan jet is marked with a white rectangle. a) Continuum intensity 4000\,\AA. b) \cair line core intensity. c) Ca II K line core intensity d) \halpha \: line core intensity. e) Enlarged view of white rectangle in panel d. A red, blue, green and orange dot show the physical locations from which the profiles shown in panel f were taken. f) \halpha \: profiles of flare ribbon (red) and fan jet (blue, green, yellow).  }
   \label{fig:overview}
\end{figure*}

In Fig.~\ref{fig:overview} we give an overview of the field of view (FoV). In the 4000~\AA \: continuum (panel a) we see a large delta sunspot crossed by a light bridge a (x,y) = 12", 20" arcsec.
In the chromospheric lines (panel $b-d$) we see a pair of large flare ribbons belonging to the X9.3 flare, that appear bright in all three panels, which expand outwards over time. In front of the eastern flare ribbon (left), at (x,y) = 15", 20" arcsec a fan jet rooted above the aforementioned light bridge appears as a dark feature against the bright flare ribbon in all the chromospheric filters. This structure is most clearly seen in the \halpha $\:$ line. We have marked the region containing this feature with a white dotted rectangle in each of the frames and enlarged it in panel $e$. This area is the FoV subsequently used in this article to study the jet. With an apparent maximum extent of roughly 6 Mm in \cair this is a very small example of such a jet compared with those typically studied in the literature. We arrive at the estimate of 6 Mm using the apparent deprojected locations of the footpoint of the structure seen in the coaligned \cak 4000 \AA ~continuum point and the head of the jet observed in \cair.

In order to summarise the pattern of \halpha \: profiles observed in this region, four locations have been selected from the data at points shown by the colored dots in panel $e$. The corresponding \halpha \: profiles are shown in panel $f$, using the same colors. The background flare ribbon (red), contains spectral profiles that have similar intensities across the entire wavelength window. This results from line broadening in flares that has been observed to extend up to 5 \AA \: from the line center \citep{Ichomoto1984, wuelserr1989, zarro1988}. Highly broadened profiles in this flare were also presented in \citet{2020Zharkov} and \citet{2020Zharakova}. Profiles taken from locations where the fan jet is visible (blue, green and yellow) show background levels consistent with the flare ribbon emission in the far wings, and with absorption features caused by the material in the fan jet elsewhere. Profiles taken further towards the center of the fan generally show broader absorption profiles with more flattened cores. In this paper we focus primarily on the \halpha \: data, using the \cair only to obtain an estimate of the fan temperature. 

\begin{figure}
   \centering
   \includegraphics[width=\columnwidth, trim=0cm 0cm 0cm 0cm,clip]{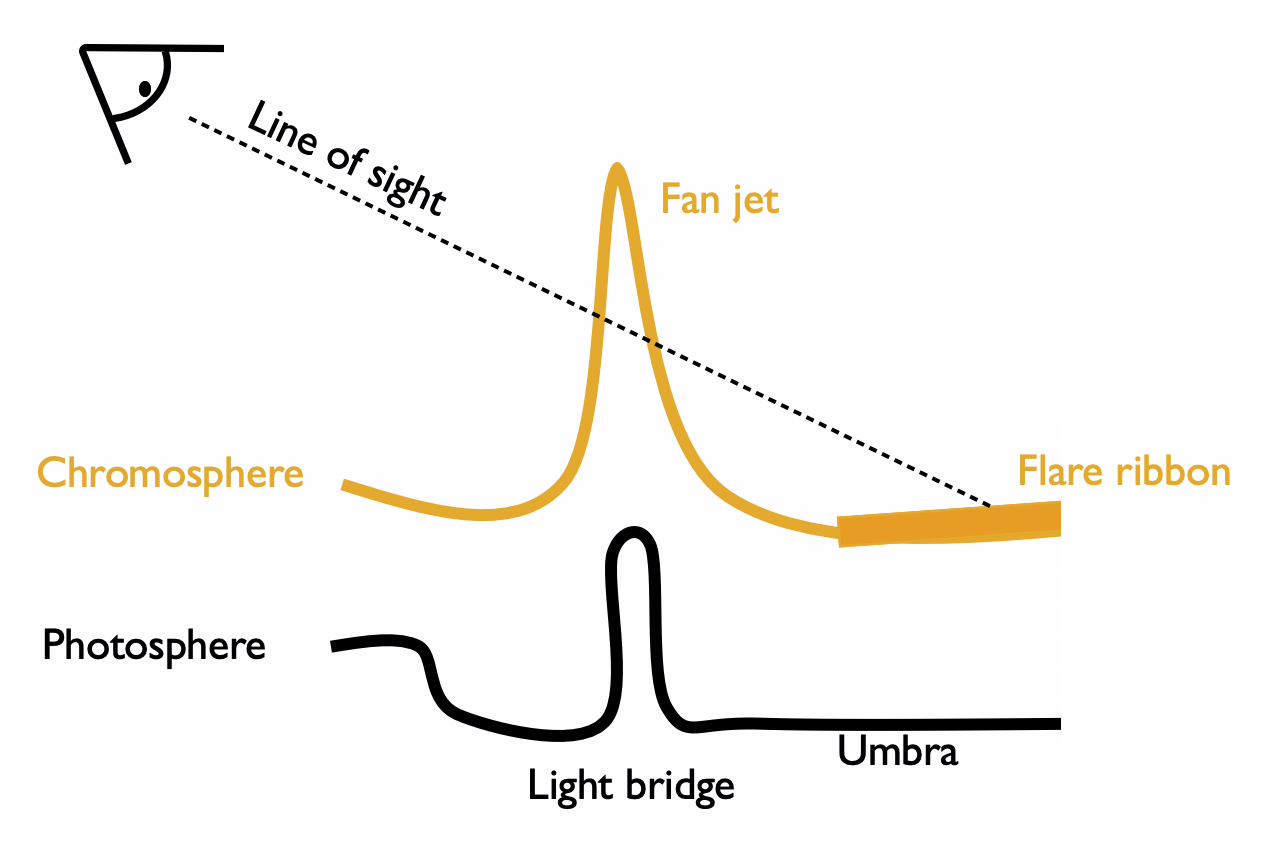}
   \caption{A cartoon showing a simplified side-on view of the physical situation being studied at two optical depth values for the photosphere and chromosphere. A fan jet is situated above a light bridge with a hot flare ribbon behind it. From our perspective the fan is backlit by the ribbon.}
   \label{fig:cartoon}
\end{figure}

\begin{figure*}[!htp]
   \centering
   \includegraphics[width=\textwidth, trim=4cm 0cm 3cm 0cm,clip]{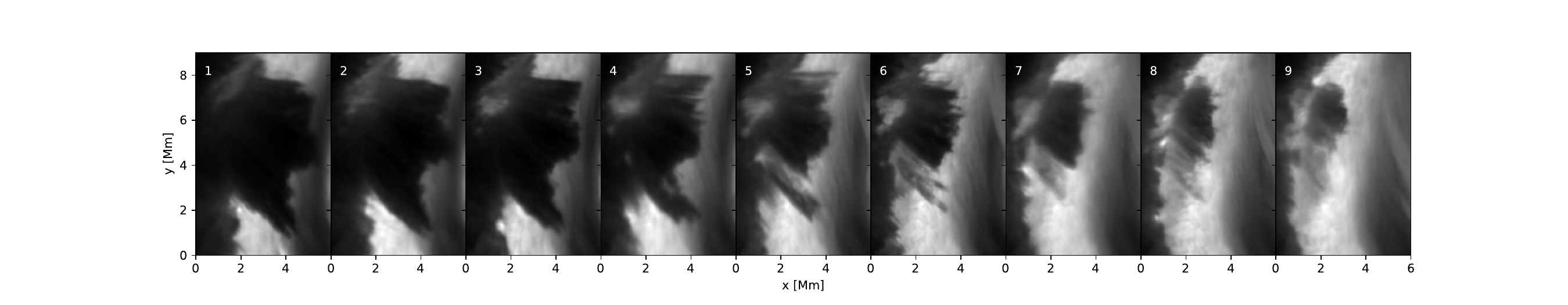}
   \caption{Timeseries of fan jet from 11:59 UT to 12:01 UT with 15~s between each panel.}
   \label{fig:timeseries}
\end{figure*}

As the flare ribbon expands, it reaches the base of the fan jet (at the location of the light bridge) and the fan begins collapsing. This suggests that the flare ribbon is suppressing the flows of material upwards from its base. The collapse of the fan jet is shown in Fig. \ref{fig:timeseries}, between 11:59 and 12:01 UT. The images in the time series are consistent with the jet being oriented approximately perpendicularly to the solar surface. This statement is based on comparing the angle that would be derived from the observing angle ($37^\circ $) to the viewing angles calculated by using line-of-sight and plane-of-sky velocities of fan material (which had a mean value of $ 44^\circ $, over the "spines" of the fan that are described in sect.\ref{velocity}). This study focuses only on the collapse of the fan jet to estimate its mass, due to poorer seeing in the time frames before the collapse began. From the LOS velocities and the visible top of the jet, we infer that the jet collapses from the second frame of Fig. \ref{fig:timeseries}, since its structure remains near identical between the first two frames. This suggests that we have reasonable seeing from the start of the collapse of the fan in our limited time series. A qualitative representation of the physical situation of the observations can be seen in Fig. \ref{fig:cartoon}.

\section{Data Inversions}\label{datainversions}
\subsection{Beckers' cloud model}\label{BCM}
We performed inversions of the \halpha \: profiles from the selected FoV by using Beckers' cloud model \citep[BCM,][]{Beckers64}. This is a solution to the radiative transfer equation for an atmosphere with a 'slab' or 'cloud' of material above it. In this model the cloud absorbs radiation coming from behind it. The optical depths can then be derived from the model by the absorption profiles that are superimposed on the background \citep{Tziotziou07}. In our model the background is the flare ribbon emission and the cloud is the material in the fan jet, see Fig.~\ref{fig:cartoon}.

In order to solve the radiative transfer equation, Beckers made several assumptions for this model. Firstly, he assumed that the slab is fully separated from the underlying atmosphere, equivalent to the assumption that the material in the fan jet is spatially separated from the material in the flare ribbon, behind it in the LOS. Secondly, he assumed that the slab is homogeneous along the line-of-sight, meaning that the inferred parameters are also constant along the line-of-sight. This assumption will not be strictly satisfied for the material in the fan jet, but should be sufficient to provide an estimate of the material in the fan jet in each pixel in the field of view. Finally, Beckers' cloud model requires the background intensity to be known. Additionally, in this unique setup of a fan jet that is back-lit by a flare, the highly broadened and flattened \halpha \: emission profiles from the flare ribbon allow us to approximate the fan jet as if it were back-lit by continuum emission (Figs.~\ref{fig:cartoon} and ~\ref{fig:overview} f, red line). One objection to this approximation would be that the flare ribbon profiles could be strongly peaked outside of the spectral window. However, due to the differences in wavelength, peaks in the full \halpha \: emission profiles that occur outside of the spectral window, like those observed in \cite[]{wuelserr1989}, will have very low probabilities of interacting with the absorption in \halpha \: resulting from the material in the fan. The assumption that the flare ribbon emissions are similar behind the fan to those in the area surrounding it is corroborated by the similar blue wing intensities (e.g. the curves in Fig. \ref{fig:overview}f at -1.5 \AA), and also the similar, flattened flare profiles seen simultaneously in all regions of the main flare ribbon around the fan.

Under these assumptions the radiative transfer equation simplifies to

\begin{equation}
    \centering
    I(\Delta \lambda) = I_0(\Delta \lambda) \: e^{-\tau (\Delta \lambda)} + S ( 1 -  e^{-\tau (\Delta \lambda)}),
    \label{eq:1}
\end{equation}
where $I(\Delta \lambda)$ is the observed intensity, $I_0(\Delta \lambda)$ the background intensity, $\tau(\Delta \lambda)$ the optical depth and $S$ the source function of the cloud material. Typically these models assume either a Gaussian or Voigt profile for the absorption profiles as a function of wavelength, which then gives a similar profile over wavelength in optical depth \citep{Tsiropula99}. For our study we selected the more general Voigt profile. Our model uses the $lmfit$ Python package \citep{lmfit} along with the Bound Limited memory Broyden–Fletcher–Goldfarb–Shanno algorithm \citep[L-BFGS-B][]{Byrd1995} to minimize the residuals between observed input profile and synthetic spectra in an iterative manner. 

\begin{figure}
   \centering
   \includegraphics[width=\columnwidth, trim=0cm 0cm 0cm 1.8cm,clip]{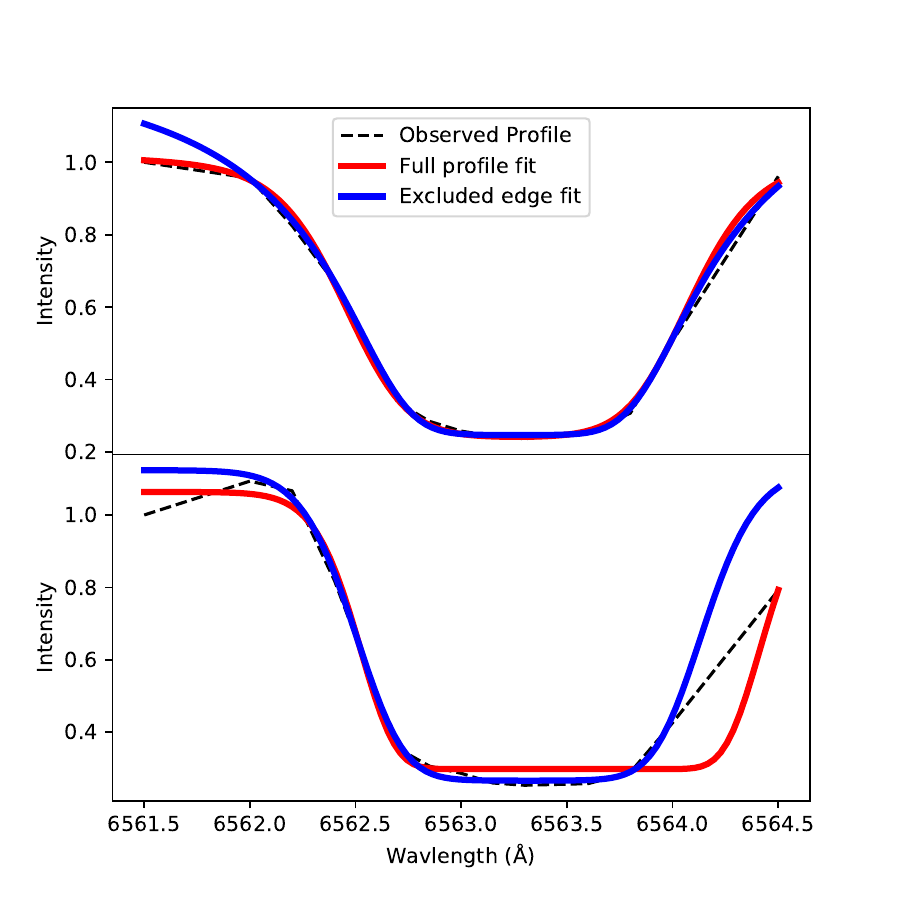}
   \caption{Plot of examples of \halpha \: profiles from pixels containing the fan jet (black) and the fit using Beckers' cloud model including (red) and excluding (blue) the two outer-most wavelength points. Top: A fitted profile with a low $\chi^2$, showing good agreement between the blue and red curves. Bottom: A profile with asymmetrical far wings that caused a high $\chi^2$ value in the original fit, and a much better fit to the main component shown by the blue curve.}
   \label{fig:level12}
\end{figure}

The code treats each pixel separately, that is to say that their corresponding profiles are fitted independently from each other. A second fit was made to the same profiles excluding the outer-most wavelength points, in order to obtain better fits to the profiles with stronger asymmetries in the far wings. These values are used to replace results from the first fit where the residuals were greater than 0.8. We illustrate the difference between the two fits in Fig. \ref{fig:level12}, where we see a symmetrical profile that the cloud model easily reproduces the profile in its entirety, while a profile that has asymmetrical far wings was more dubiously fitted. Neglecting the two outer-most points allowed us to accurately fit the main curve of the absorption profile, and hence provide a more conservative estimate of the optical depth for the main component of the material in the fan jet (see Fig \ref{fig:level12}, lower panel, blue line).

In order to select the pixels that we were confident to contain material in the fan jet, we selected only profiles with an absorption feature. This was achieved by requiring the quotient of the lowest intensity across all the wavelength points to the intensity in the central wavelength point to be 0.8 or higher. The resulting mask of selected pixel was then manually checked, and anomalous pixels were removed to form a final mask for each time frame. This means that an area smaller than that of the full fan is used during our inversions, resulting in an influence towards underestimation of the total optical depth and mass contained.

Using this model we inverted our \halpha \: time series, fitting for the following parameters: the optical depth of the line core $\tau_0$, the Doppler shift, the damping parameter of the Voigt function $\Gamma$, the Gaussian width of the Voigt function $\sigma$ (which includes Doppler broadening, microturbulence and other effects), $I_0$ and $S_0$. All values except the Doppler shift were bounded to be positive. $I_0$ and $S_0$ are assumed to be constant along the wavelength range, with their initial guesses inferred from the profile minimum and maximum. 

To reproduce the widths of these profiles a lot of broadening was required. This implies some combination of high Gaussian broadening (i.e. increased $\sigma$ via high temperatures or microturbulent velocities) and additional effects such as Lorentzian broadening, velocity gradients within the formation region, or optically thick line formation effects, such as opacity broadening (see \citet{molnar2019}).

The study of \citet{2009cauzzi} using cloud modelling for \halpha \: required adjustments equivalent to temperatures of up to 60kK to fit their profiles. Additionally, \citet{Tziotziou072} found that the \halpha \: line widths in quiet Sun observations were strongly affected by the seeing and by the fact that the retrieved optical depth values are negatively skewed, once again resulting in an influence towards some large underestimations of optical depths and therefore the total mass contained in the fan jet.

Modelling the broadening principally by using higher $\tau_0$ values does not result in good fits of the full profiles. That is to say "opacity broadening"  principally results in saturation of the intensity profile in the core to the source function values of the cloud at high opacity, meaning that a broader profile will also have a flatter core. This saturation happens for a greater span of the wavelengths in the absorption profile as $\tau_0$ increases significantly above unity. However, this fitting via higher optical depth is particularly poor in the line wings, which become much steeper than in the observations and thus result in large values of $\chi^2$. The vast majority of the profiles are symmetrical and thus the widths of these profiles are unlikely to be the result of strong velocity gradients. If they were, this would imply that very particular patterns of velocity gradients that happen to result in symmetrical profiles are present in the majority of pixel, across the field of view and for different angles of orientation between the fan spines and the viewer. There are some small fraction of profiles present that do show some asymmetries that would be better explained using this information, but these profiles have a main absorption feature in line with the widths of profiles in surrounding features, and an additional element that likely represents a feature created by a strong velocity gradient. In the small number of more asymmetrical profiles our routine focuses on fitting what appears to be the main absorption feature of the profile, e.g. see Fig. \ref{fig:level12}. The best fits produced a close match to both the absorption core and the gradients of the wings. This was achieved by the algorithm in almost every case by using lower but still optically thick $\tau$ values coupled with substantial broadening of the Gaussian width parameter $\sigma$. Asymmetric profile features  resulting from velocity gradients are commonly observed in flares but can also be found in quiet sun features as well, and have been investigated in numerous studies such as \citet{2015Kuridze, 2017Druett, 2017Capparelli, 2018Druett}. For example, in our data we see a "horn" of the \halpha ~profile in the blue wing or a lower gradient away from the central Doppler shifted component in the red wing, which are both visible in lower panel of Fig.~\ref{fig:level12}, black line.

This high $\sigma$ value for \halpha~ remains unexplained and may imply high turbulent broadening or temperatures. This is certainly worthy of detailed investigation that is not within the remit of this paper. We infer temperature by other, more reliable methods using a different spectral line (see sect.\ref{sec:temp}) and use this cloud model to focus only on the $\tau_0$ and Doppler shift. We conjecture that due to expected temperature estimates obtained via other spectral lines, that if $\sigma$ is indeed large, there must be significant microturbulent broadening in \halpha. However, due to the non-degenerate $\chi^2$ values for our fits of the \halpha~ line, we remain reasonably confident of the estimates of $\tau_0$ inferred. This kind of large scale \halpha~ broadening is also present in flares, and is a still unexplained phenomena that we took advantage of by assuming a flat profile of continuum-like back-lighting.

Once we had an estimate of $\tau_0$ for each pixel in our fan, we converted this optical depth into a column mass using the relation found in \citet{jorrit2012}, in which the authors showed that the \halpha \: line optical depth loses sensitivity to variations of temperature under chromospheric conditions (i.e. variations within typical chromospheric temperatures, that are sufficiently above that of the temperature minimum region, but below the ionisation temperature of hydrogen). The opacity becomes mainly determined by the mass density, which in turn implies that the optical depth is proportional to the column mass. We use this proportionality in our work to convert the optical depth directly into a column mass ($m \: \mathrm{[g\ cm^{-3}]} = 3\times 10^{-5} \times \tau_0 $). The column mass per pixel is equal to the total observed mass in the fan jet per pixel in our case, as the absorption is caused by the cloud only.

If we assume a uniform thickness for the fan in the line of sight, it is also possible to translate the mass estimate into a density estimate, by dividing the column mass by the thickness of the fan jet in the line of sight. A thickness estimate of the fan was obtained by looking at the narrowest width of an individual fan structure that was resolved in the images, and estimated at around 200 km, which matches the typical width of a narrow fibrilar structure \citep{sepideh20, Zhou2020}.  Then density figures can be compared our values to similar phenomena and potentially to simulations at a later stage. We can use this density estimate to create a second cloud model that fits the fan's \cair profiles, as described in the following section. 

This choice for thickness is likely to be an underestimate, as it supposes that the narrowest observed feature is suitable for use as the general estimate for the thickness of the fan. There is a direct relationship between the thickness and the resulting densities that are derived from our estimate of the mass, namely that we could choose a larger thickness and thus get lower densities. However, in the following section we found that our model is not able to reproduce the \cair ~spectral profile as successfully when using a thickness for the slab that is larger by a factor greater than two, without also requiring higher density values. We highlight to the reader that the assumptions made in calculating the mass estimate of the fan jet lead to a result that is a lower limit, whereas the subsequent assumption made for the thickness of the fan influences the estimates of densities towards greater values and thus cannot be considered formally as either an upper or lower limit.

\subsection{Temperature fitting}\label{sec:temp}
In order to estimate the temperatures of the flare and the fan jet (and thus to corroborate our prior assumption that the temperature in the fan jet is chromospheric at the beginning of the collapse) we constructed a second cloud model. This model employed the STockholm inversion Code (STiC) to perform inversions of the reduced \cair \: profiles from the selected FoV. The profiles in \cair from the fan jet and nearby flare ribbon were not broadened greatly by unexplained processes at the time of the fan collapse (unlike in \halpha) and are also resolved within the wavelength window of the observations. Therefore, it would not have been reliable to use Becker's cloud model with an approximation of a flat continuum background in \cair, nor would it be reliable to fit any other background profile without taking into account spatial dependence. In the method described below we restricted ourselves to comparing the flare and fan jet information in spatially close locations, near to the edge of the jet to maximise the likelihood of similarity in the background emission profiles from the flare (see Fig.\ref{fig:results2}c). 

The STockholm inversion Code\footnote{\url{https://github.com/jaimedelacruz/stic}} \citep[STiC;][]{Jaime16,Jaime19} is an MPI-parallel NLTE inversion code that utilises a modified version of RH \citep{Uitenbroek01} to solve the atom population densities assuming statistical equilibrium and plane-parallel geometry. The radiative transport equation is solved using cubic Bezier solvers \citep{jaime13}. The inversion engine of STiC includes an equation of state extracted from the SME code \citep{2017piskunov}. STiC uses a regularized version of the Levenberg-Marquardt algorithm \citep{levenberg44, marquardt63} to iteratively minimize the $\chi^2$ function between observed input data and synthetic spectra of one or more lines simultaneously. The code treats each pixel separately as a plane-parallel atmosphere, fitting them independently of each other. 

Using STiC we inverted representative \cair profiles of the flare ribbon to obtain a model atmosphere for the background\footnote{We used an inversion process similar to the one discussed in section 4.3 of \citet{pietrow2020}.}. A uniform slab, representing the fan jet, with a thickness of 200 km was inserted at the upper boundary of the atmosphere, above the flare ribbon atmosphere derived from the first step. We chose to put the slab on top because the background atmosphere is in hydrostatic equilibrium, and because the slab is introduced at the upper boundary, in a density and temperature regime where the Ca II 8542 is otherwise not sensitive, it does not affect the background atmosphere.  Unlike the slab model used to estimate the mass density, this slab is introduced in a depth-stratified atmosphere, by setting the upper $200$~km of the atmospheric model with constant values of $T, v_{\mathrm{los}}, v_{\mathrm{turb}}$ and the estimated mass density. The slab gently transitions to the pre-existing parameter stratification in a three grid points after the first $200$~km. This is done for numerical reasons, because when STiC solves the NLTE problem, changes in line-of-sight velocity between grid points that are larger than half the Doppler width of the line will induce errors in the intensities.

To estimate the values of the parameters in the slab, STiC is only used as a forward modeling synthesis tool to compare the resulting profile with the observations, and the inversion is done by the Levenberg-Marquardt algorithm. We do not consider hydrostatic equilibrium in this experiment, and thus turned it off for these inversions.

The profile from the fan jet was fitted using the parameter space of the slab (T, $v_{\mathrm{los}}$ and $v_{\mathrm{turb}}$) via a Levenberg-Marquardt minimization routine\footnote{\url{https://github.com/jaimedelacruz/LevMar}} \citep{Jaime19}, while keeping the rest of the model stratification fixed, in a way similar to the one described in \citet{Carlos19}.  

We repeated this inversion, varying the slab thickness with values from 100 km up to 800 km, and proportionally scaled densities. For values of the slab thickness from 200 km up to 400 km, there was a degeneracy between the density and thickness, and we were able to achieve similar quality fits with these degenerate values. However, the final $\chi^2$ values of the fits was higher for larger values for the fan thickness, and more dramatically so above 400 km and below 200 km. For this reason, we continue our investigation by assuming a slab thickness of 200 km.

Regardless of these variations, we derived similar slab temperatures, all within the uncertainty in sect. \ref{sec:restemp}, for each of the inversions that were able to fit the observed profile. Therefore, we believe that while there may be some degeneracy in our inferred densities and slab thickness, a robust value for the temperature of the slab is derived.

\begin{figure*}[!htp]
   \centering
   \includegraphics[width=\textwidth, trim=0cm 0cm 0cm 0cm,clip]{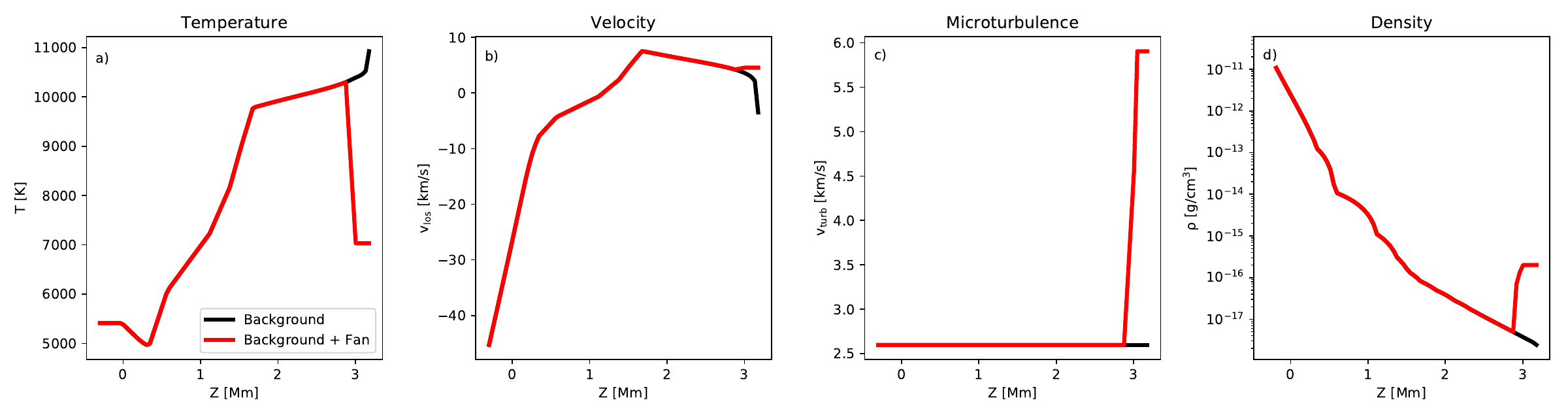}
   \caption{Stratification in geometrical height scale of parameters ($T, v_{\mathrm{los}}, v_{\mathrm{turb}}$) inverted with our second cloud model (described in sect. \ref{sec:temp}) as well as the inferred density from the first cloud model (described in sect. \ref{BCM}). The black line shows the stratification of the aforementioned parameters of flare ribbon background, while the red line compares these same parameters for the best fit of an inverted pixel inside of the fan. Note that the model stratification is fixed everywhere except in the slab at the top of the atmosphere. In panel a we show the temperature, in panel b the line-of-sight velocity, in panel c the microtubulence and in panel d the density. The observed and best-fitted profiles belonging to these atmospheres can be found in Fig. \ref{fig:results2}c.}
   \label{fig:stratification}
\end{figure*}

\subsection{Collapse velocities and acceleration}\label{velocity}
\begin{figure}
   \centering
   \includegraphics[width=\columnwidth, trim=2cm 0cm 2cm 1.2cm,clip]{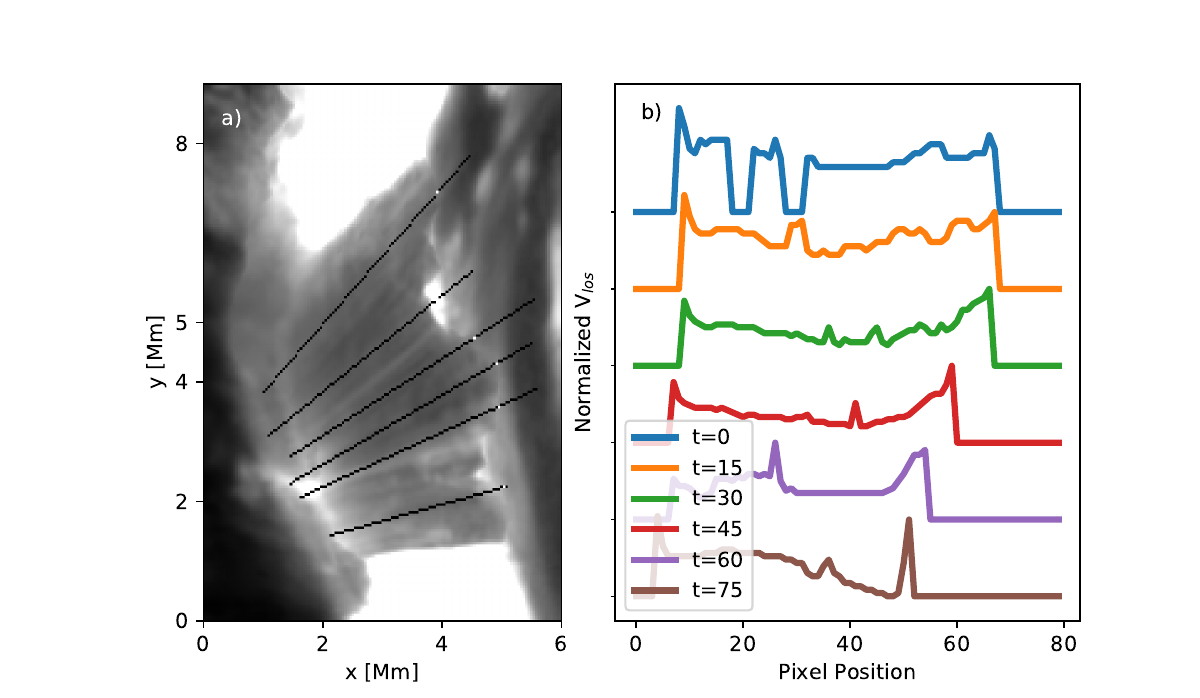}
   \caption{The collapse of the fan was tracked along six 'spines'. Panel a) An  of the minimum value of the \halpha~ emission intensity. This shows the fan in the first time frame (11:59:20 UT). The six spines used are shown using black lines, with a white dot denoting the edge of the fan as derived from the process shown in panel b. b) shows the velocity values on the y-axis (normalised to the same maximum value for comparison) against position along the length of the spine in pixels on the x-axis (zero being the base of the spine). This data is shown for the fourth spine from the top at six different times in different colors. We chose this spine due to its central location inside of the fan, this location had a clear dark appearance and was less susceptible to ionization over the time frames considered.} 
   \label{fig:vlostrack}
\end{figure}
We obtained velocity estimates for the collapse of the fan jet by combining measurements of the line-of-sight and plane-of-sky velocities. The latter was found by tracking the top of the fan jet along 6 trajectories, or fan 'spines', which were aligned with visible structure within the fan (See Fig. \ref{fig:vlostrack}a) during the first 6 frames of the collapse. The values became unreliable for the later frames of the time series because large parts of the fan lost visibility in \halpha, likely due to energy from radiation and heating that is ionizing the neutral hydrogen (see Fig. \ref{fig:timeseries}. The top of the fan on each spine is found from the fact that the BCM breaks down on the edge, at this point there ceased to by an clear absorption profile and the fitting routine attempts to interpret the resulting asymmetric line profile via a single Voigtian. This generally produced a high Doppler shift causing a strong peak in the line-of-sight velocity parameter at the point where, moving upward along the spine you reach the top of the fan, before dropping to 0~km~s$^{-1}$ outside of it, because the emission (flare ribbon) or absorption (off ribbon) line profiles are generally centred on the rest wavelength. This can be seen on the right side of each profile in Fig. \ref{fig:vlostrack}b, and the collapsing top of the fan can be tracked along this spine by noting the change in position of this peak over time. We illustrate the collapse of the fourth spine in Fig. \ref{fig:vlostrack}b, by plotting the $v_{\mathrm{los}}$ along this spine in the first six time frames. The collapse can be inferred from the position of the peak in LOS velocity on the right side of each profile in this panel. 

This method ignores any curvature of the fan spines, and sizeable uncertainties remain due to the seeing quality. Therefore, only the six clearest spines were used in order to minimise the uncertainties arising from the ionisation of material or cooling of material falling from above. Spines in the outer regions of the fan seem to be particularly difficult to track, possibly because of their greater variations in angle of ejection, movements, and ionisation of the material in the jet appearing to start earlier in these regions.

The line-of-sight velocity was fitted from derived Doppler velocities of BCM model resulting from the \halpha \: profiles. However, as can be seen in Fig. \ref{fig:vlostrack}b, the fitting procedure breaks down towards the edge of the fan, giving a much higher velocity in the last few pixels. Therefore, we evaluated the velocity in the more reliable internal region a few pixels below the top of the spines. 

Acceleration was then derived from the time derivatives of the velocities.

\subsection{Momentum and Energy Estimate}\label{momentum}
Using the mass and velocity estimates, we could in principle calculate the momenta, and thus the momentum delivered to the base of the fan jet as a function of time by looking at the mass difference between each time frame. However, while this is one of the main parameters that are used to describe classical sunquake amplitudes \citep[e.g.][]{2020Sharykin,2020Stefan}, these momenta are typically the integrated total of what is delivered to the surface over the full lifetime of the event, and this lifetime is unfortunately seldom reported. However, one such duration of a momentum delivery period has been reported for this flare to have a duration of around 40s \cite{2020Zharakova}. In contrast to the flaring events that drive sunquakes, fan jets have been known to last for tens of minutes up to a hour, therefore we do not believe that citing a total momentum would be a meaningful comparison in this case, as this would result in momenta that are several orders of magnitude larger than for reported sunquakes. The same can be argued for the total energy of the quake. A more suitable number to use as a comparison is the energy density flux of the event, which can be found by dividing the peak kinetic energy per second of the fan that is delivered to the solar surface by its area\footnote{Which we estimate to be 5.5Mm $\times$ 200 km = $1\times10^{16}$ cm$^2$.}. 

The aforementioned ionization of the fan prevents us from investigating the full time series of the collapse due to the "disappearance" of the material in the \halpha\: observations, which would result in an overestimated momentum/energy value. To mitigate this, we used the acceleration derived from the previous section and combined it with the second frame\footnote{We chose the second frame because it gives the highest mass.} in our time series to model a collapse of the material over time. The mass distribution was calculated for the fan in the second frame on a per pixel basis. Each pixel was given an initial velocity based on the assumption of falling from the top of the fan under the acceleration derived from the previous step. Each pixel was also assigned a distance from the spine foot point based on an orthogonal projection onto the nearest spine, based on the six spines from Fig. \ref{fig:vlostrack}a. Thus, we estimated the momentum and energy delivery to the base of the fan jet per second. 

\begin{figure*}
   \centering
   \includegraphics[width=\textwidth, trim=3.5cm 0cm 5.5cm 0cm,clip]{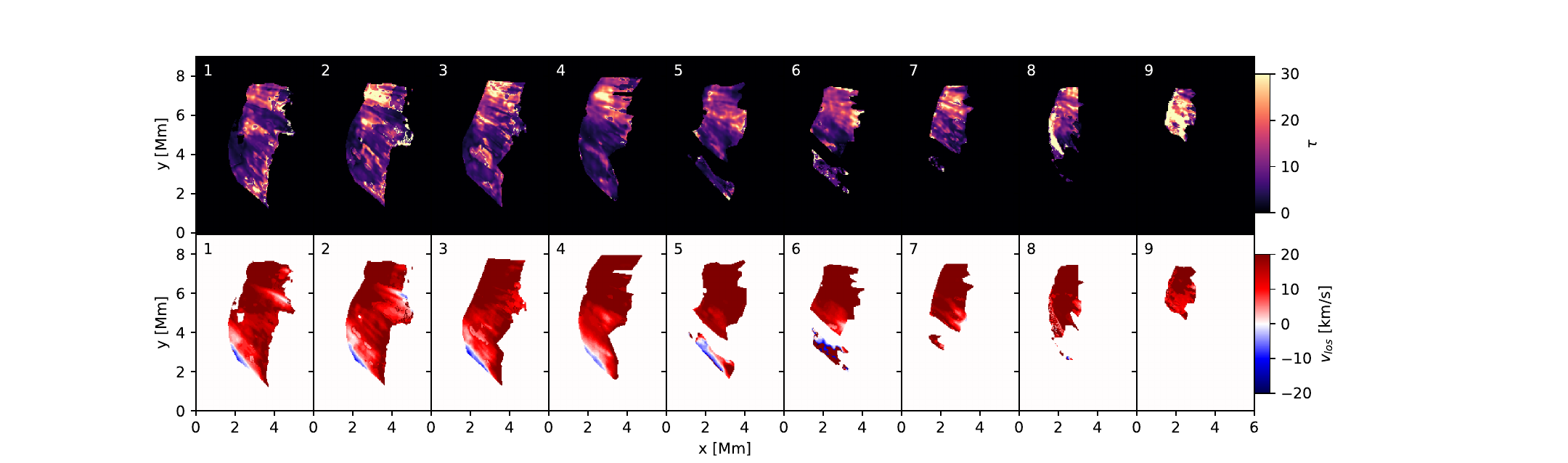}
   \caption{Best fit results of BCM inversions of time series presented in Fig. \ref{fig:timeseries}. Top row: Optical depth Bottom row: v$_{\mathrm{los}}$.}
   \label{fig:results1}
\end{figure*}

\begin{figure*}
   \centering
   \includegraphics[width=\textwidth, trim=3cm 0cm 3.5cm 0cm,clip]{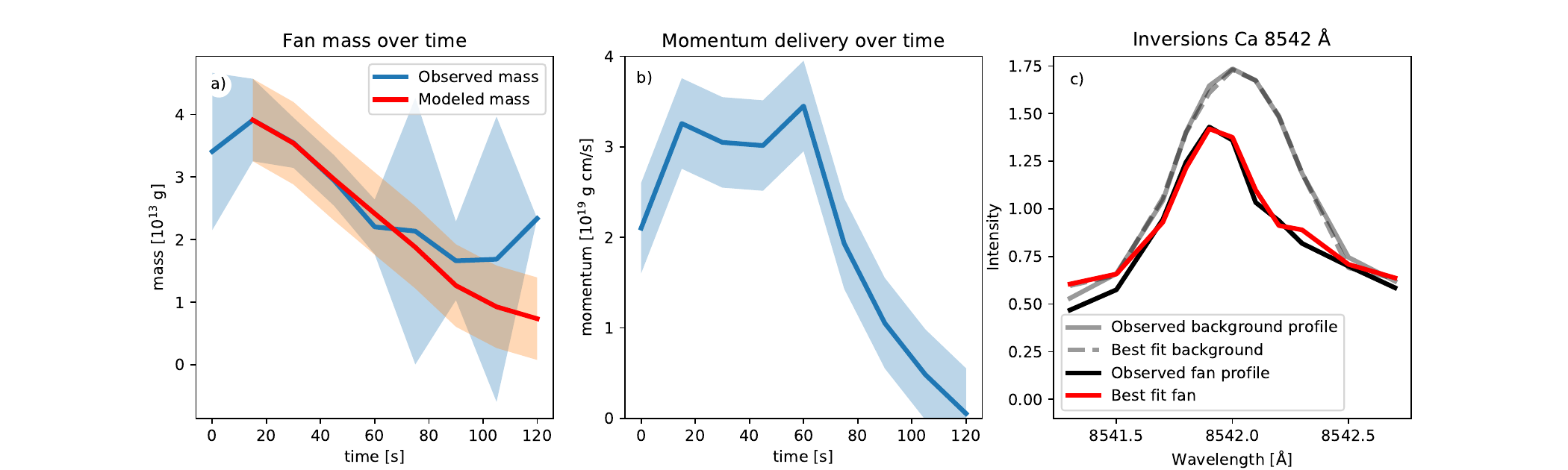}
   \caption{a) Observed total mass of jet over time (blue). Simulated fan collapse based on 2nd time frame assuming acceleration under gravity (red). b)  Delivered momentum to solar surface over time. c) \cair \: STiC inversion of flare background (gray dashed), obtained from observed profile (gray). \cair \: fan profile (black) fitted from background profile with an absorbing slab above it. }
   \label{fig:results2}
\end{figure*}

\subsection{Uncertainties}
We obtain the uncertainties for the estimates of the velocity and optical depth from our first cloud model via the $lmfit$ python package, which works similarly to the optimization library of SciPy, but including uncertainties and correlations between fitted variables \citep{lmfit}. We obtain our uncertainty on the temperature inversion from the second cloud model by varying the parameters for the cloud until the model until the resulting profile no longer fitted the observed data. Classical error propagation theory was used for the total velocity estimate, as well as for the value of the mass, momentum and energy density flux. 

\section{Results and Discussion}\label{resultsndisc}

In Fig.~\ref{fig:results1}, we present the BCM inversion results for the line core optical depth and line of sight velocities of nine time frames of the $6\arcsec \times 13\arcsec$ field of view that was marked with a white dotted line in Fig.~\ref{fig:overview}. Pixels outside of the fan have been masked as described in section \ref{BCM}. 

\subsection{Optical depth}
The optical depth maps in the top row of Fig. \ref{fig:results1} show that the overall structure of the fan is reasonably uniform. The inversions for most of the pixels give a $\tau_0$ value close to 8 (purple), with some higher opacity regions where the $\tau_0$ can reach around 20. In regions where the cloud model assumptions break down, sometimes anomalous values of $\tau_0$, greater than 100, are returned. This is mainly due to the fact that the variations in the profile background become comparatively large to the slope of the wing or shape of the core. This made the code prioritize fitting variations in broadened flare background profiles over the overall shape of the absorption profile. Greater values are typically found at the top of the fan and were removed by the mask. The very top of the fan is a region in which some of the BCM model assumptions become less accurate. For example, when it is not true that the fan is back lit in the line-of-sight by the same spectrally broad flare ribbon emission (See Fig. \ref{fig:timeseries}1-3). 

The model also breaks down in cases where the optical depth is too low for the core of the line profiles to saturate and thus flatten (e.g. the blue line in Fig. \ref{fig:overview}f). This case leads to an overestimation of the source function, which results in a higher fitted optical depth \citep{diaz2018analysis}. In other works this problem was typically solved by taking $S = \frac{1}{2}I$ (e.g. \citep{1997Paletou, 2006heinzel, 2010labrosse}), to avoid the degeneracy between $\tau_0$ and $S$ \citep{diaz2018analysis}. However, in our case we know $S$ for pixels where the fan material gives a saturated line profile and only face this degeneracy for a small number of pixels where the fan material is optically thin. 

For this reason we have capped the value for $\tau_0$ at 30. This preserved all the reliable results derived from inversions of pixels within the main body of the fan, while avoiding anomalous and unreliable results from a few pixels dominating the mass estimate. Later in the time sequence this issue becomes more pronounced due to the heating of material in the fan which results in a dramatic increase in the number of inversions with low saturation that had much larger uncertainties. Because these problems became increasingly acute with time, we restricted our analysis to the subset of the time sequence for which the results were reliable. 

\subsection{Velocity of the collapse}
The derived LOS velocity, shown in the lower row of Fig. \ref{fig:results1}, displays higher values at the base of the fan jet and for a greater fraction of the mask as time progresses, in line with the expected behaviour as the fan collapses. It also shows generally higher values in LOS velocity and $\tau$ towards the top of the field of view, strongly suggesting that the inclination of the fan to the observer is greater in these locations, this is corroborated further by the fact that LOS velocities shrink towards zero or are slightly oriented towards the observer (blue shifted) at the very bottom of the FOV in the mask.

The POS collapse of the fan has been tracked as explained in section \ref{velocity} along six spines that are illustrated in Fig. \ref{fig:vlostrack}a. The fitted acceleration in this plane was found to be $a_{POS} = 0.29 \pm 0.19 $~km s$^{-2}$. When combined with the values found in the LOS, we get a resultant acceleration of $a_{abs} = 0.39 \pm 0.12 $ km s$^{-2}$. This is within the error bars for acceleration under solar gravity and other forces not playing a significant role, which is also consistent with the findings of \citet{2012morton} and \citet{carolina16}, but not with those of \citet{1973roy} and \citet{2018Reid} where a decelerating force is reported. A possible explanation for the absence of this deceleration is that the processes that cause the upward jets of the fan are suppressed by the flare ribbon as well as the fact that there is no upwards-moving fan material in the same plane as the downwards-moving fan material. As mentioned earlier, a small signature of the base of the fan remained after the main collapse, which could alternatively be interpreted as a deceleration of the original collapse near the base. This can be seen in the blue line of observed mass in Fig. \ref{fig:results2}a at times after 60s.

However the uncertainties in these results are quite large, and also increase dramatically after frame 5, due to issues such as ionisation of the plasma, and so we are not able to complete a full study of the collapse using solely the observational data. Therefore, in our model of the complete collapse of the fan we use the mass distribution that was derived observationally from the first frames, and assume that the structure collapsed under gravity, as was suggested by the frames that are most reliable for the estimation of the acceleration in the collapse. The maximum absolute velocity of falling material at the bottom of the fan is 70 km s$^-1$.

\subsection{Fan mass}
Using the relations shown in section \ref{BCM} we can convert our reported optical depth values into masses, and subsequently use an estimate of the geometrical thickness of the fan to convert the masses into densities. The total mass for each of the nine time-frames shown in Fig. \ref{fig:timeseries} is plotted in blue in Fig. \ref{fig:results2}a, with the peak value of $3.9 \pm 0.7 \times 10^{13}$ g being found in the second time frame, at the start of the collapse. This significantly smaller than the value of $10^{15}$ g suggested by \citet{1973roy} for a large surge, based on the assumption of an isotropic expansion of an initial density of $3 \times 10^{15}$ particles cm$^{-3}$ from the base of the surge. This discrepancy could be due to more simple assumptions used in Roy's estimate, because our surge is smaller, or a natural variation of the physical parameters of such jets. However, the height of the fan observed in Roy's work was 10 times greater than that reported here. Multiplying the mass estimate for our fan by a factor of 10 to account for this height difference produces the mass estimates that are only separated by a factor of 3.

The similar masses in the first and second frame, as well as the nearly identical shape of the fan in these two frames, as seen in Fig. \ref{fig:timeseries}, suggests that the collapse does not start until the second time frame. From the second to the fifth frame we get a smooth collapse, and after which the observed total mass starts to fluctuate. This is also the time frame when the ionization of the fan begins to dominate the opacity changes on significant fractions of the total area of the fan jet. However, we note that the collapse seems to halt as the fan persists for a number of frames at a much smaller size of around 2 Mm.

The total masses derived from the collapse described in section \ref{velocity} is overplotted in red in Fig. \ref{fig:results2}a, with error estimates shown using the lighter shading. This collapse was modelled using the mass distribution from the second frame and gravitational acceleration, starting at the 2nd time frame. We can see that it matches the observed collapse (blue curve) very well for the first three time steps (15 - 60 s), after which the values derived from the two methods diverge and large error estimates affect the observational values. From this data it seems that there is still some smaller vertical range near the base over which the peacock jet was acting and interacting with the flare ribbon during this time, but that the material higher continued collapsing under gravity. Thus, the picture is less clear after 60s in the data.

\subsection{Fan density}
Using the aforementioned assumption that the fan has a uniform thickness of 200 km (based on the width of the thinnest resolved structures in the plane-of-sky direction), we also report estimated mean densities. For the convenience of the reader we provide this number in three different unit formats, $2 \pm 0.3\times10^{-8}$ kg m$^{-3}$, $2 \pm 0.3\times10^{-11}$ g cm$^{-3}$ or $9 \pm 1.4 \times10^{12}$ particles cm$^{-3}$. This is a typical value for chromospheric material and within the same order of magnitude of the value of $10^{12}$ particles cm$^{-3}$ suggested by \citet{1973roy} for a large surge, based on the assumption of an isotropic expansion of an initial density of $3 \times 10^{15}$ particles cm$^{-3}$ from the base of the surge. Our value is also consistent with values for the simulated chromospheric jets reported by \citet{2013Takasao}. \citet{2013Kayshap} reported a much lower value of 4.1 $\times$ 10$^9$ particles cm$^{-3}$, but this is for a much hotter surge of 2MK. We can also compare our densities to similar chromospheric structures, for example a Bifrost fibril has a typical density of around $10^{-9}$ to 10$^{-10}$ kg m$^{-3}$ \citep{2022druett} and for spicules densities around 
$2.2 \times 10^{10}$ particles cm$^{-3}$ have been reported \citep{2000Sterling, 2020Shimojo}.

\subsection{Fan temperature} \label{sec:restemp}
Our second slab model used the STiC code, and allowed us to fit a fan profile\footnote{The fitted fan profile was taken at around [x,y]=5,7 Mm on Fig. \ref{fig:overview}1.}  from the \cair \: observations. This was achieved by using a profile from the nearby flare background and a slab with the thickness and density that were reported above, as is discussed in section \ref{sec:temp}. In Fig. \ref{fig:results2}c we can see a \cair \: flare background profile in gray and a fan profile in black. The dashed gray curve represents the STiC inversion made of the profile, which was then used as the input model for our fan fitting routine. The red curve was obtained from this input model after only modifying the temperature, LOS velocity and microturbulence of the cloud part of the model. The best fitted temperature for the background was between 10 and 11 kK at chromospheric heights, which matches findings from \citet{Rahul21}. For the slab material the best fitted temperature was around $7050 \pm 250$ K. The model also produced estimates for the slab material of LOS velocity (4.5 km s$^{-1}$) and microturbulent velocity (6 km s$-1$). The LOS velocity is comparable to the value found from the BCM model for this pixel of 5.5 $\pm$ 1.4 km s$^{-1}$, 
which increases our confidence in the temperature estimate, which is typical for chromospheric material, as was required by our assumptions for the relation between $\tau_0$ and column mass from \citet{jorrit2012}. 

\citet{carolina16} reported an average brightness temperature (the temperature of a blackbody that would produce this intensity in the wavelength of the \halpha~ line core of 5.5 kK that was consistent  over several different pixels for a similar jet observed in \halpha. This is lower than our NLTE inversion value, but matches the \halpha \: brightness temperature calculated using the same pixel that was employed for our inversions by looking at the \halpha ~line core intensity as described in \citet{carolina16}. We stress the point that although these results are consistent, since both structures are NLTE we should not assume that this brightness temperature value is a true temperature of the plasma or conclude that the NLTE plasma temperatures of the fan discussed here and the one reported in \citet{carolina16} match. Our inverted temperature is lower than simulated surge temperatures of $10^5$~K  or higher as reported by \citet{2008Nishizuka} and \citet{2013Kayshap}, however such temperatures have been contested by \citet{2016nobrega} who suggest a typical value around $10^4$~K when radiative losses are considered. 

We note that the \halpha~ profile widths of the fan jet material in this dataset are much larger than those reported in \citet{molnar2019}. Extrapolating the best fit line from this work would suggest temperatures of up to 25kK for the \halpha~ profiles presented in our work. This is far beyond the ionization temperature of \halpha. The majority of broadening in the work of \citet{molnar2019} was attributed to opacity effects that happened to correlate with changes in ALMA brightness temperature. The relationship between the line width and plasma temperatures inferred in \citet{molnar2019} appears to break down in the context presented here. 

\subsection{Fan momentum and energy}
It is possible to deduce an estimate of the momentum delivered at the base of the jet as a function on time by using the model of the collapse of the fan as described in section \ref{momentum}. The results of this estimate are shown in Fig. \ref{fig:results2}b. Most notable here is the near-constant values between the 2nd and 5th time frame, which peaks at 3.8 $\pm$ 0.5 $\times 10^{18}$ g cm s$^{-1}$ for a 1 second interval. This is at least two orders of magnitude smaller than most of the reported sunquakes \citet{2020Stefan}, including the ones found in our dataset. However, as discussed before it is hard to compare to this number without knowing the time over which the momentum has been delivered. If we multiply the number by 40~s, to make the times comparable to the sunquake reported in \citet{2020Zharakova}, we end up a factor of 40 below the lowest total momentum of $5.4\times10^{21}$ g cm s$^{-1}$ that was fitted to a sunquake in our dataset by \citet{2020Stefan}. This suggests that our fan jet does not supply sufficient energy to act as the source of a sunquake beneath it. However, if we artificially scale up our fan model to 200~Mm by extending the mass distribution linearly in height, we obtain values with a similar order of magnitude, which opens up the possibility for a large fan jet is an energetically plausible source for triggering a sunquake. 

A similar calculation provides an estimate of the energy density flux instead of the momentum. This gives us a value of $1.0~\times~10^{9}$~erg~s$^{-1}$~cm$^{-2}$, which is three orders of magnitude lower than the lower estimate of $1.5 \times 10^{12}$ erg s$^{-1}$ cm$^{-2}$ reported by \citet{2017Sharykin}. \citet{2017Sharykin} reports on a sunquake that has a total momentum only one order of magnitude higher than the smallest momentum detected for sunquake to date \cite{2020Stefan}. In terms of energy density flux we can only reach 10\% of the reported value by scaling up the fan to 200~Mm. This suggests that while a similar total momentum flux could be achieved in a larger fan, the area over which the energy is spread might be too large to trigger a visible sunquake.

\section{Conclusions}\label{conclusions}
We have investigated a small fan jet above a light bridge of a sunspot in AR12673. The upward jet was suppressed by a flare ribbon that expanded over its base during an X9.3 flare. This ribbon covered the sunspot umbra and prevented the subsequent ejection of material up from above the light bridge. This created a unique situation where the remaining material of the fan could be studied while it was falling. Our main focus was on the \halpha \: observations of this dataset, aided by the fact that the flare ribbon profiles had flattened profiles with similar intensities along the entire CRISP wavelength window. This allowed us to employ a simple cloud model to perform inversions that could fit the profiles and deduce estimates of the optical depth of the material in the collapsing jet.

This simple model produced good quality  fits to the profiles for most of the pixels in the fan over the course of nine time frames (See Fig. \ref{fig:results1}) and thus to provide estimates of the optical depths and LOS velocities of the material in the fan for each pixel. These optical depths could then be used to estimate the masses of material in each pixel of the fan using the relation from \citet{jorrit2012}. The peak value for the total mass in the jet came in the 2nd time frame, with a value of $3.9 \pm 0.7 \times 10^{13}$ g and a density of $2 \pm 0.3\times10^{-11}$ g cm$^{-3}$. 

Absolute velocity estimates could be found by combining the LOS velocities in each pixel with the POS velocities obtained from tracking the top edge the fan during the first five time frames of the collapse. The resulting estimates of the acceleration were consistent with solar gravity. 

A temperature of $7050 \pm 250$ K was found for a sample of material in the fan using a second, independent cloud model based on the synthesis mode of the STiC code and an external inversion tool. Here we fitted a \cair\ fan profile by first inverting a flare ribbon profile, and using this atmosphere  as a background atmosphere  for the fitting of the fan profile with an additional slab placed at the upper boundary of the background atmosphere. The fitting routine then estimated the fan profile by only modifying the parameters of the slab (see Fig. \ref{fig:results2}c). This temperature is lower than the one found in literature for similar structures. However, the 5500~K brightness temperature of material in our fan jet matches the value found in \citet{carolina16}. Our temperature is chromospheric as was required by our assumptions for the relation between $\tau_0$ and column mass from \citet{jorrit2012}. 

In order to achieve an estimate of the momentum delivered to the base of the fan jet, we first constructed a model of the fan collapse. This was because effects such as ionization made some parts of the fan transparent during the later stages of its collapse. This made estimates of the acceleration derived from observations unreliable in the later time frames. Moreover, some small jet material remained visible at the base for a while longer. The red curve in Fig. \ref{fig:results2}a shows values based on the collapse of the initial mass distribution of the fan under gravity,  without these effects. From this model of the collapse it was possible to make an estimate of the  momentum delivery to the base of the fan against time, which is shown in Fig. \ref{fig:results2}b. The peak value of this momentum delivery was 3.5 $\pm$ 0.5 $\times 10^{19}$ g cm s$^{-1}$. This is two orders or magnitude smaller than the momentum delivery reported in most sunquakes \citep{2020Stefan}, including some found in this dataset \citep{2020Zharakova, 2020Zharkov}. However we do remark that these momenta are for the entire event and our value is given as a delivery per second. A more meaningful comparison is given by looking at the energy density flux, which gives us a value of $1.0 \times 10^{9}$ erg s$^{-1}$ cm$^{-2}$, which is three orders of magnitude lower than a reported value by \citet{2017Sharykin} for a larger sunquake. Therefore, the collapse of this small fan jet can be ruled out as the source of a sunquake in this observation. However, since this is a very small jet, this result does not completely exclude other, larger jets from being energetically plausible triggers for sunquakes. Our estimates do suggest, however, that very large jets of 200 Mm or more would be needed to reach values comparable to those that have been observed to date.

Our results were fundamentally enabled by the unique nature of this dataset, as we took advantage of a very specific physical set up. Therefore, we cannot easily compare our findings to other datasets with more variable background emission. Thus, a good next step would be to compare our finding to simulated peacock jets. Additionally there are other unique sets of conditions that could yield a similar simplification (such as limb observations). Additionally, one could search for sunquake signatures under larger jets and, if found, attempt to derive estimates of the momentum delivery from falling material in the fan jet in this manner.

\begin{acknowledgements}
MD was supported by The Swedish Research Council, grant number 2017-04099.
JdlCR gratefully acknowledges funding from the European Research Council (ERC) under the European Union's Horizon 2020 research and innovation program (SUNMAG, grant agreement 759548).
The Swedish 1- m Solar Telescope is operated on the island of La Palma by the Institute for Solar Physics of Stockholm University in the Spanish Observatorio del Roque de los Muchachos of the Instituto de Astrof\'isica de Canarias. The Institute for Solar Physics was supported by a grant for research infrastructures of national importance from the Swedish Research Council (registration number 2017-00625).
Computations were performed on resources provided by the Swedish Infrastructure for Computing (SNIC) at the PDC Centre for High Performance Computing (Beskow, PDC-HPC), at the Royal Institute of Technology in Stockholm as well as the National Supercomputer Centre (Tetralith, NSC) at Link\"oping University.
This work was supported by the Knut and Alice Wallenberg Foundation.
This research has made use of NASA's Astrophysics Data System Bibliographic Services. 
We acknowledge the community effort devoted to the development of the following open-source packages that were used in this work: numpy (numpy.org), matplotlib (matplotlib.org), astropy (astropy.org).
We thank Adur Pastor Yabar and Carlos Diaz Baso for their valuable conversations and suggestions.
We thank Mihalis Mathioudakis, Sean Quinn, and Aaron Reid for providing access to the Queen's University Belfast observational campaign data sets from the SST. UK access to the Swedish 1-m Solar Telescope was funded by the Science and Technology Facilities Council (STFC) under grant No. ST/P007198/1

\end{acknowledgements}

\bibliographystyle{aa}
\bibliography{ref}

 \begin{appendix}

\end{appendix}
\end{document}